\documentclass{article}
\usepackage[T2A]{fontenc}
\usepackage[cp866]{inputenc}
\usepackage[tbtags]{amsmath}
\usepackage{amsfonts,amssymb,mathrsfs,amscd,comment}

\begin{document}

\title{\textbf{{\large \MakeUppercase{On quantum Gaussian optimizers
conjecture in the case} $q=p$}}}
\author{A.S.~Holevo \\ Steklov Mathematical Institute, RAS}
\date{}
\maketitle

\makeatletter \renewcommand{\@makefnmark}{} \makeatother \footnotetext{%
The work was supported by the grant of Russian Scientific Foundation
(project No 14-21-00162).}

\makeatletter\renewcommand{\@makefnmark}{} \makeatother\footnotetext{}

Let $\mathcal{H}$ be a separable Hilbert space, $\mathfrak{\ T}_{p},\,1\leq
p<\infty ,$ the Schatten class i.e. the Banach space of operators  in $%
\mathcal{H}$ with the finite norm $\left\Vert X\right\Vert _{p}=\left(
\mathrm{Tr}\left\vert X\right\vert ^{p}\right) ^{1/p}$ ;  $\mathfrak{T}%
_{\infty }$ is the space of compact operators with the operator norm. For $%
1\leq p,q\,\leq \,\infty $ the norm of a linear map $\Phi :$ $\mathfrak{T}%
_{p}\rightarrow \mathfrak{T}_{q}$ is defined as
\begin{equation}
\left\Vert \Phi \right\Vert _{q\rightarrow p}=\sup_{X\neq 0}\left\Vert \Phi
\lbrack X]\right\Vert _{p}/\left\Vert X\right\Vert _{q}  \label{norm}
\end{equation}%
provided it is finite.

The \textit{quantum Gaussian optimizers conjecture}, having roots in
analysis \cite{lieb} and in quantum information theory \cite{H2}, says that
if $\Phi $ is Bosonic Gaussian channel (see below) then the supremum is
attained on \textquotedblleft Gaussian\textquotedblright\ operators $X.$ In
the papers \cite{GHM}, \cite{obzor} the conjecture was proven for $q=1$ and
all gauge-covariant (contravariant) Gaussian channels with $s$ degrees of
freedom. For arbitrary $p,q>1$ the result was obtained in \cite{PTG} for
rather special class of gauge-covariant channels with $s=1$. Recently the
hypothesis was confirmed (in the asymptotic sense) for $q=p$ and
gauge-covariant channels with $s\geq 1$ \cite{FL}. The proof is based on a
general upper bound from \cite{beigi} and results from our paper \cite{H1}
concerning Gaussian channels. In the present note we show that the results
of \cite{H1} and \cite{HSH} in fact allow to prove the hypothesis for $q=p$
and \textit{all} Gaussian channels. Thus the condition of gauge covariance,
rather crucial in the case $q=1,s>1,$ plays no role in the case $q=p.$

Let $(Z,\Delta )$ be the standard symplectic vector space ($\dim Z=2s$) with
the nondegenerate skew-symmetric matrix $\Delta $, and let $W(z)=\exp
(iRz);\,z\in Z$ be the Weyl system in a Hilbert space $\mathcal{H}$ giving a
regular representation for the Canonical Commutation Relations.
Here $R$ is the $2s$-vector row of the canonical observables. Gaussian density operator (d.o.)
$\rho $ with mean $m$ and the covariance matrix $\alpha $ satisfying $\alpha
\pm \frac{i}{2}\Delta \geq 0$ is defined by the quantum characteristic
function
\begin{equation}
\mathrm{Tr}\rho W(z)=\exp \left( im^{\top }z-\frac{1}{2}z^{\top }\alpha
z\right) .  \label{CF}
\end{equation}%
Let $\Phi $ be a general Bosonic Gaussian channel i.e. completely positive
map $\mathfrak{T}_{1}\rightarrow \mathfrak{T}_{1}$ such that
\begin{equation}
\Phi ^{\ast }\left( W(z)\right) =W(Kz)\exp \left( il^{\top }z-\frac{1}{2}%
z^{\top }\mu z\right) ,  \label{BGC}
\end{equation}%
where real symmetric matrix $\mu $ satisfies $\mu \geq \pm \frac{i}{2}
\left( \Delta -K^{\top }\Delta K\right).$ The channel $\Phi $ transforms a
Gaussian d.o. with the covariance matrix $\alpha $ into Gaussian d.o. with
the covariance matrix
\begin{equation}
\alpha ^{\prime }=K^{\top }\alpha K+\mu ,  \label{CM}
\end{equation}%
see \cite{H2} for more detailed description.

\textsc{Theorem.} \textit{Let $\Phi $ be a Bosonic Gaussian channel (\ref%
{BGC}) with invertible matrix $K$, then
\begin{equation}
\left\Vert \Phi \right\Vert _{p\rightarrow p}=|\det K|^{1/p-1},\quad  1\leq p \leq\infty , \label{npp}
\end{equation}
with the supremum in (\ref{norm}) attained in the limit on a sequence of Gaussian d.o. If $q<p,$ then $\Phi $ does not extend to a bounded map $\mathfrak{T}%
_{p}\rightarrow \mathfrak{T}_{q}$.}

\textsc{Proof.} Let us first prove the upper bound
\begin{equation}
\left\Vert \Phi \right\Vert _{p\rightarrow p}\leq |\det K|^{1/p-1}.
\label{upper}
\end{equation}%
The general interpolation bound of \cite{beigi} implies, as in \cite{FL},
that for a positive map $\Phi $
\begin{equation*}
\left\Vert \Phi \right\Vert _{p\rightarrow p}\leq \left\Vert \Phi
\right\Vert _{\infty \rightarrow \infty }^{1-1/p}\,\left\Vert \Phi
\right\Vert _{1\rightarrow 1}^{1/p},
\end{equation*}%
where $\left\Vert \Phi \right\Vert _{\infty \rightarrow \infty }=\left\Vert
\Phi \lbrack I]\right\Vert $ (in the infinite dimensional case $\Phi \lbrack
I]$ should be defined as in  \cite{H1}) and $\left\Vert \Phi \right\Vert
_{1\rightarrow 1}=1$ in the case of channel.

Now let the matrix $K$ be invertible, then it is proven in \cite{H1} that
\begin{equation}
\Phi \lbrack I]=|\det K|^{-1}I,  \label{PI}
\end{equation}
hence (\ref{upper}) follows.

To prove asymptotic achievability of (\ref{upper}) we follow \cite{H1}, \cite%
{HSH}. We assume $p<\infty ,$ the case  $p=\infty $ can be treated
similarly. In \cite{HSH} it was proven (Eq. (28)) that
\begin{equation}
\mathrm{Tr}\rho ^{p}W(z)=\left[ \det \,f_{p}(\mathrm{abs}(\Delta ^{-1}\alpha
_{\beta }))\right] ^{-\frac{1}{2}}\exp \left[ im^{\top }z-\frac{1}{2}z^{\top
}{\alpha }g_{p}\left( \mathrm{abs}(\Delta ^{-1}{\alpha })\right) z\right] ,
\label{chara1}
\end{equation}%
where $f_{p}(d)=(d+1/2)^{p}-(d-1/2)^{p}\sim p\,d^{p-1}$ as $d\rightarrow
\infty .$ For $z=0$ we obtain%
\begin{equation*}
\mathrm{Tr}\rho ^{p}=\left[ \det \,f_{p}(\mathrm{abs}(\Delta ^{-1}\alpha
_{\beta }))\right] ^{-\frac{1}{2}}.
\end{equation*}%
Let $\rho =\rho _{\beta }$ be the d.o. of the Gibbs state with
the Hamiltonian $H=R\epsilon R^{\top },$ where $\epsilon $ is a
nondegenerate positive definite matrix (e.g. unit matrix), with the inverse
temperature $\beta $. Then $\rho _{\beta }$ is Gaussian with the covariance
matrix \cite{H1}%
\begin{equation}
2\Delta ^{-1}\alpha _{\beta }=\cot \beta \epsilon \Delta .  \label{cot1}
\end{equation}%
Consider the asymptotic $\beta \rightarrow 0$. Then from (\ref{cot1}) $%
\alpha _{\beta }\sim \left( {2\beta \epsilon }\right) ^{-1}$ and from (\ref%
{CM}) $\det \alpha _{\beta }^{\prime }\sim |\det K|^{2}\det \alpha _{\beta }.
$ Hence%
\begin{equation*}
\mathrm{Tr}\rho _{\beta }^{p}\sim p^{-s}\left[ \det \,(\mathrm{abs}(\Delta
^{-1}\alpha _{\beta }))\right] ^{\frac{1-p}{2}}=p^{-s}\left\vert \det
\,(\Delta ^{-1}\alpha _{\beta })\right\vert ^{\frac{1-p}{2}%
}=p^{-s}\left\vert \det \,\alpha _{\beta }\right\vert ^{\frac{1-p}{2}}.
\end{equation*}%
Similarly
\begin{equation*}
\mathrm{Tr}\,\Phi \lbrack \rho _{\beta }]^{p}\sim p^{-s}\left\vert \det
\,\alpha _{\beta }^{\prime }\right\vert ^{\frac{1-p}{2}}\sim p^{-s}|\left(
\det K\right) ^{2}\det \,\alpha _{\beta }|^{\frac{1-p}{2}}\sim \left\vert
\det K\right\vert ^{1-p}\mathrm{Tr}\rho _{\beta }^{p}.
\end{equation*}%
Hence $\lim_{\beta \rightarrow 0}\frac{\mathrm{Tr}\,\Phi \lbrack \rho
_{\beta }]^{p}}{\mathrm{Tr}\rho _{\beta }^{p}}=\left\vert \det K\right\vert
^{1-p}$ and taking into account (\ref{upper}), equality (\ref{npp}) follows.

The second statement follows from the fact that $\left\Vert \rho _{\beta
}\right\Vert _{p}$ scales as $\beta ^{s\frac{p-1}{p}}$ while $\left\Vert
\Phi \lbrack \rho _{\beta }]\right\Vert _{q}$ as $\beta ^{s\frac{q-1}{q}}$
so that the ratio of the norms goes to $\infty $ for $q<p.$


\end{document}